\documentclass[twocolumn,journal]{IEEEtran}
\usepackage{graphicx}
\usepackage{amsmath}
\usepackage{amssymb}
\usepackage{tabu}
\usepackage{comment}
\usepackage{multicol}
\usepackage{amsthm}
\usepackage{cite}
\usepackage{float}
\usepackage{color}
\usepackage{balance}

\begin{document}
\title{\LARGE Physical-Layer Security of a Buffer-Aided \\ Full-Duplex Relaying~System \vspace{-0.2\baselineskip}}
 \author{Ahmed El Shafie$^\dagger$, Ahmed Sultan$^\star$, Naofal Al-Dhahir$^\dagger$
\begin{tabular}{c}
\small $^\dagger$University of Texas at Dallas, USA. \\
\small $^\star$King Abdullah University of Science and Technology (KAUST), Saudi Arabia.
\end{tabular}
\thanks{This paper was made possible by NPRP grant number 6-149-2-058 from the Qatar National Research Fund (a member of Qatar Foundation). The statements made herein are solely the responsibility of the authors.}
\thanks{This paper is published in IEEE Communications Letters.} \vspace{-2\baselineskip}}
\date{}
\maketitle
\begin{abstract}
This letter proposes a novel hybrid half-/full-duplex relaying scheme to enhance the relay channel security. A source node (Alice) communicates with her destination node (Bob) in the presence of a buffer-aided full-duplex relay node (Rooney) and a potential eavesdropper (Eve). Rooney adopts two different relaying strategies, namely randomize-and-forward and decode-and-forward relaying strategies, to improve the security of the legitimate system. In the first relaying strategy, Rooney uses a codebook different from that used at Alice. In the second relaying strategy, Rooney and Alice use the same codebooks. In addition, Rooney switches between half-duplex and full-duplex modes to further enhance the security of the legitimate system. The numerical results demonstrate that our proposed scheme achieves a significant average secrecy end-to-end throughput improvement relative to the conventional bufferless full-duplex relaying scheme.
\end{abstract}
\section{Introduction}
 The use of buffers to enhance secrecy rates has been investigated recently in, e.g., \cite{huang2015buffer,7296606}. In \cite{huang2015buffer}, Huang and Swindlehurst investigated the physical (PHY) layer security of a buffer-aided half-duplex (HD) relay channel. The authors proposed an adaptive link selection scheme for time slot assignments between the source and relay nodes. However, the proposed scheme is suboptimal in the sense that it ignores the impact of the buffer state information at the relay node. Reference \cite{huang2015buffer} assumed that there is {\bf no} direct link between the source node and both the legitimate destination node and the potential eavesdropping node. Unlike \cite{huang2015buffer}, we consider a \emph{buffer-aided full-duplex (FD) relay with finite buffer}. In \cite{7296606}, the authors proposed two secure cooperative schemes for a wireless relay channel with buffered relay and source nodes. In contrast to \cite{7296606}, we assume an FD relay node with a reliable power supply.

 Recently, the authors of \cite{chen2015physical} investigated the PHY-layer security of a single-input single-output single-antenna eavesdropper (SISOSE) wiretap channel with \emph{bufferless FD single-antenna relay node}. Similar to \cite{huang2015buffer},  Reference \cite{chen2015physical} assumed that there is no direct link between the source node and its destination. In \cite{7500135}, the authors proposed three relay selection
schemes for FD heterogeneous networks in the presence
of multiple cognitive radio eavesdroppers. The relay selection schemes are based the availability of the
network's channel state information (CSI) at the secondary information
source.

The contributions of this letter are summarized as follows
 \begin{itemize}
       \item We propose a novel hybrid HD/FD relaying scheme and two decoding strategies at the relay node to enhance the security of the legitimate system. Our proposed scheme exploits the \emph{exact} buffer state information at the relay node and the secrecy rates of all links.
           \item We derive the system secrecy rate and average end-to-end secure throughput. Moreover, we convert the throughput-maximization problem into a standard linear program.
              \end{itemize}
\textit{Notation}: $\mathbf{I}_{\mathcal{B}}$ denotes the identity matrix whose size is $\mathcal{B}\times \mathcal{B}$. $\mathbb{C}^{M \times N}$ denotes the set of all complex matrices of size $M\times N$ and $\mathbb{C}$ denotes the set of all complex numbers. $(\cdot)^*$ denotes Hermitian operation.
\section{Main Assumptions}
We consider a wireless network composed of one source node (Alice), one eavesdropping node (Eve), one FD relay node (Rooney), and one destination node (Bob). All nodes are equipped with a single antenna \cite{chen2015physical}. Alice is always backlogged with data packets to transmit. We assume that the distance between Alice and Bob is large enough due to shadowing and large distances such that the direct link between them is completely disconnected as in, e.g., \cite{huang2015buffer,chen2015physical}. Eve eavesdrops both Alice's and Rooney's transmissions. Time is slotted into discrete equal-size time slots. The duration of a time slot is $T$ seconds and the channel bandwidth is $W$ Hz.

We assume fixed secrecy-rate transmissions since in real systems the data is packetized and arrives in packets from the upper layers. Therefore, in a given time slot, the legitimate transmitting nodes transmit fixed-length packets of size $b_{\rm s}$ bits over one time slot duration. Since the transmission time is $T$ seconds and the bandwidth is $W$ Hz, the {\bf transmission secrecy rate} is $\mathcal{R}_{\rm s}=\frac{b_{\rm s}}{W T}$ bits/sec/Hz.
\vspace{-0.25cm}
\subsection{Relay's Model}
\vspace{-0.05cm}
Rooney has an FD capability and is equipped with a finite data buffer to store Alice's data. The buffer can store up to $\mathcal{Q}$ packets. The number of packets in the buffer is $Q_R$, $0\le Q_R\le \mathcal{Q}$. If Rooney transmits data from his buffer, he employs a randomize-and-forward (RF) relaying strategy where he uses a codebook different from that used by Alice. The advantage of this relaying strategy is that Eve cannot combine the data transmitted by Alice and Rooney. Thus, the achievable secrecy rate is enhanced \cite{mo2012relay}. If Rooney's buffer is empty, while it is still feasible to operate in the FD mode, Alice and Rooney transmit the data cooperatively using a secrecy rate that is based on the minimum of the two-hop transmission rates. The transmitted packet is delivered to Bob directly without being stored in Rooney's buffer.

In summary, we have three main transmission schemes: 1) RF-FD where the FD capability is utilized in the RF mode; 2) DF-FD where the FD capability is utilized in the DF mode; and 3) HD mode where either Alice or Rooney transmits when FD transmissions are insecure.
\subsection{Channel Model}
Each link experiences block flat-fading where a channel coefficient remains unchanged during one time slot duration and changes independently from one time slot to another. We denote the channel coefficient between Node $n_1$ and Node $n_2$ in time slot $t$ by $h_{n_1n_2}(t)$, where $n_1,n_2\in\{A,B,E,R\}$ denote Alice, Bob, Eve, and Rooney, respectively. The complex random variable $h_{n_1n_2}(t)$ is independent and identically distributed (i.i.d.) from one time slot to another with zero mean and variance $\sigma_{n_1n_2}^2$. Each link is also corrupted by an additive white Gaussian noise (AWGN) process with zero mean and variance $\kappa_{m}$ Watts/Hz. In a given time slot, all nodes are assumed to have complete and perfect CSI \cite{chen2015physical}. This is reasonable when all nodes are active (non-hostile). The CSI is obtained through channel estimation and feedback. It is assumed that the average transmit power at Node $n_1$ is $\mathcal{P}_{n_1}$ Watts \cite{huang2015buffer}.
\subsection{Received Signal-to-Interference-Plus-Noise Ratios}
In this subsection, we analyze the received signal-to-interference-plus-noise ratios (SINRs) at all nodes.
\subsubsection{FD mode}
The received SINR at Rooney is given by
\begin{equation} \small
\begin{split} \small
\label{cocowawa2}
\gamma_{R,{\rm FD}}(t)= \frac{|h_{AR}(t)|^2\mathcal{P}_A}{|h_{RR}(t)|^2 \mathcal{P}_R+\kappa_R W}
\end{split}
\end{equation}
where $h_{RR}(t)\in \mathbb{C}$ is the residual self-interference at Rooney after interference cancelation. The term $|h_{RR}(t)|^2 \mathcal{P}_R$ in the denominator of \eqref{cocowawa2} represents the self-interference power. Thus, the received SINR at Rooney is degraded due to self-interference caused by the FD operation, which, in turn, decreases the secrecy rate of the Alice-Rooney link. This demonstrates the benefit of HD modes when the self-interference power is high. The received SINR at Bob is given by
\begin{equation} \small
\begin{split} \small
\label{cocowawa2xx}
\!\!\!\!\!\gamma_{RB}(t)\!=\! \frac{|h_{RB}(t)|^2 \mathcal{P}_R}{\kappa_B W}
\end{split}
\end{equation}

In the RF-FD mode, the data signal transmitted by Alice, $x_A(t)$, and the data signal transmitted by Rooney, $x_R(t)$, are two independent signals. In the DF-FD mode, we assume $x_R(t)=x_A(t-1)$.\footnote{The data processing delay is assumed to be one data symbol.} The latter case resembles an inter-symbol interference (ISI) channel.
\subsubsection{HD mode}
 If Rooney operates in the HD mode, $x_R(t)~=~0$. Assuming Alice only transmits, the SINR at Rooney is given by
 \begin{equation} \small
\begin{split} \small
\label{cocowawa2x}
\gamma_{AR,{\rm HD}}(t)= \frac{|h_{AR}(t)|^2\mathcal{P}_A}{\kappa_R W}
\end{split}
\end{equation}
 In addition, the received SINR at Eve is thus
\begin{equation} \small
\begin{split} \small
\label{cocowawa2xx}
\!\!\gamma_{AE,{\rm HD}}(t)\!=\! \frac{|h_{AE}(t)|^2\mathcal{P}_A}{\kappa_E W}
\end{split}
\end{equation}
If Rooney only transmits, the SINR at both Bob and Eve are
 \begin{equation} \small
\begin{split} \small
\label{cocowawa2x}
\gamma_{RB,{\rm HD}}(t)= \frac{|h_{RB}(t)|^2\mathcal{P}_R}{\kappa_B W}, \
\!\!\gamma_{RE,{\rm HD}}(t)\!=\! \frac{|h_{RE}(t)|^2\mathcal{P}_R}{\kappa_R W}
\end{split}
\end{equation}
\section{Secrecy Rates}
\vspace{-0.05cm}
In this section, we investigate the nodes' achievable data rates (without secrecy constraints) and the legitimate links secrecy rates. The achievable rate at Rooney is
\begin{equation} \small
\begin{split} \small
\label{relay_rate}
\!\!\mathcal{R}_{AR,{\rm M}}(t)\!=\! \log_2\left(1+\gamma_{R,{\rm M}}(t)\right)
\end{split}
\end{equation}
where ${\rm M}\in\{{\rm HD},{\rm FD}\}$ denotes the duplex mode at Rooney. On the other hand, the achievable rate at Bob is
\begin{equation} \small
\begin{split} \small \small
\label{RelayBobrate}
\mathcal{R}_{RB}(t)= \log_2\left(1+\gamma_{B}(t)\right)
\end{split}
\end{equation}
Hereinafter, we omit the time index from the rate~expressions.
\subsection{FD Mode}
In the RF-FD relay mode, the signals received by Eve from Alice and Rooney are independent. The secrecy rate region is obtained as follows
\begin{enumerate}
\item If Eve assumes Rooney's transmission as noise and only decodes Alice's transmission, the SINR and the achievable rate at Eve are given, respectively, by
    \begin{equation} \small
\begin{split} \small
\!\!\!\!\!\gamma_{AE,{\rm FD}}\!&=\!\frac{|h_{AE}(t)|^2 \mathcal{P}_A}{|h_{RE}(t)|^2 \mathcal{P}_R+\kappa_E W},\ \mathcal{R}_{AE}\!=\! \log_2\left(1\!+\!\gamma_{AE,{\rm FD}}\right)
\end{split}
\end{equation}
Hence, the secrecy rate of Alice is upper-bounded as
\begin{equation} \small
\begin{split}
\label{condition1}
 \mathcal{R}_{AR}^{\rm sec,FD}\le \left[ \mathcal{R}_{AR,{\rm FD}}- \mathcal{R}_{AE}\right]^+
\end{split}
\end{equation}
where $\left[\cdot\right]^+=\max(\cdot,0)$.
\item If Eve assumes Alice's transmission as noise and only decodes Rooney's transmission, the SINR and achievable rate at Eve are given, respectively, by
     \begin{equation} \small
\begin{split} \small
\!\!\!\!\gamma_{RE,{\rm FD}}\!&=\!\frac{|h_{RE}(t)|^2 \mathcal{P}_R}{|h_{AE}(t)|^2 \mathcal{P}_A\!+\!\kappa_E W}, \ \mathcal{R}_{RE} \!=\! \log_2\left(1\!+\!\gamma_{RE,{\rm FD}}\!\right)
\end{split}
\end{equation}
Hence, the secrecy rate of Rooney is upper-bounded as
\begin{equation} \small
\begin{split}
\label{condition2}
\mathcal{R}_{RB}^{\rm sec,FD}\le \left[ \mathcal{R}_{RB}-  \mathcal{R}_{RE}\right]^+
\end{split}
\end{equation}
\item If Eve employs a joint-typicality receiver, she can decode a sum rate
\begin{equation} \small
\begin{split}
\!\mathcal{R}_{E,{\rm sum}}\!=\!\log_2 \left(1\!+\!\frac{|h_{\rm AE}|^2  {\mathcal{P}_A}\!+\!|h_{\rm RE}|^2  {\mathcal{P}_R}}{\kappa_E W}\right)
    \end{split}
\end{equation}
 Hence, the sum secrecy rate is upper-bounded as \cite{4529293}
\begin{eqnarray}
\label{condition3}
\!\!\!\!\! \mathcal{R}_{AR}^{\rm sec,FD}\!+\!\mathcal{R}_{RB}^{\rm sec,FD}& \!\le \!\Bigg[\!  \mathcal{R}_{AR,{\rm FD}}\!+\!\mathcal{R}_{RB} \!-\!\mathcal{R}_{E,{\rm sum}}\!\Bigg]^+
\end{eqnarray}
\end{enumerate}
The rate pair $(\mathcal{R}_{AR}^{\rm sec,FD},\mathcal{R}_{RB}^{\rm sec,FD})$ should satisfy the conditions in \eqref{condition1}, \eqref{condition2}, and \eqref{condition3}.

In the DF-FD case, and assuming one symbol duration delay, the end-to-end rate of the system is given by \cite{chen2015physical}
\begin{equation} \small
\begin{split} \small
\label{total_rate}
\mathcal{R}_{AB}= \min\{\mathcal{R}_{AR,{\rm FD}},\mathcal{R}_{RB}\}
\end{split}
\end{equation}
In addition, Eve's rate is given by \cite{chen2015physical}
\begin{equation} \small
\begin{split} \small
\label{cocowawa2v233ccc}
\mathcal{R}_{E}=\frac{1}{\mathcal{B}}\log_2 \det\left(\mathbf{I}_{\mathcal{B}}+ \frac{1}{\kappa_E W}\mathbf{H}^* \mathbf{H}\right)
\end{split}
\end{equation}
where $\mathcal{B}=\lfloor WT\rfloor$ is the codeword length and $\mathbf{H} \in \mathbb{C}^{(\mathcal{B}+1) \times \mathcal{B}}$ is a Toeplitz matrix whose first column is $[h^*_{AE},h^*_{RE},0,\dots,0]^*$. The system secrecy rate is given by
\begin{equation} \small
\begin{split} \small
\label{total_rate}
\mathcal{R}^{\rm sec,FD}_{AB}=[\mathcal{R}_{AB}-\mathcal{R}_E]^+
\end{split}
\end{equation}
\subsection{HD Mode}
In the HD case, there is only one transmission from Alice to Rooney or from Rooney to Bob. In the first case, the secrecy rate of the Alice-Rooney link is given by
\begin{equation} \small
\begin{split} \small
\label{secrelay_ratexxx}
\mathcal{R}_{AR}^{\rm sec,HD}&=[\mathcal{R}_{AR,{\rm HD}}-\mathcal{R}_{AE,{\rm HD}}]^+
\end{split}
\end{equation}
where $\mathcal{R}_{AE,{\rm HD}}=\log_2(1+\gamma_{AE,{\rm HD}}(t))$.

In the second case, a transmission occurs from Rooney to Bob. The Rooney-Bob link secrecy rate is given by
\begin{equation} \small
\begin{split} \small
\label{secrelay_rate2}
\mathcal{R}_{RB}^{\rm sec,HD}&=[\mathcal{R}_{RB}-\mathcal{R}_{RE,{\rm HD}}]^+
\end{split}
\end{equation}
where $\mathcal{R}_{RE,{\rm HD}}=\log_2(1+\frac{|h_{RE}(t)|^2 \mathcal{P}_R}{\kappa_E W})$.
\section{Proposed Scheme and Rooney's Buffer}
\subsection{Proposed Hybrid HD/FD Relaying~Scheme}
\label{sec3}
The following five binary-valued quantities will be used to decide the transmission strategy in a given time slot: 1) $\mathcal{S}_1=1[\mathcal{R}_{AR}^{\rm sec,FD}\ge \mathcal{R}_{\rm s}]$, 2) $ \mathcal{S}_2=1[\mathcal{R}_{RB}^{\rm sec,FD}\ge \mathcal{R}_{\rm s}]$, 3) $ \mathcal{S}_3=1[\mathcal{R}_{AB}^{\rm sec,FD}\ge \mathcal{R}_{\rm s}]$, 4) $\mathcal{S}_4=1[\mathcal{R}_{AR}^{\rm sec,HD}\ge \mathcal{R}_{\rm s}]$, and 5) $\mathcal{S}_5=1[\mathcal{R}_{RB}^{\rm sec,HD}\ge \mathcal{R}_{\rm s}]$. The five quantities are functions of the links CSI (or the links secrecy rates). The RF-FD mode is used when $\mathcal{S}_1=\mathcal{S}_2=1$. Otherwise, it is not used. Hence, we can use $\mathcal{S}^\star=\mathcal{S}_1 \mathcal{S}_2 \in \{0,1\}$ to represent the feasibility of using the RF-FD mode.

 Our proposed scheme is summarized as~follows. When Rooney's buffer is not empty (i.e. $Q_R>0$), the legitimate system has three transmission options: 1) Alice and Rooney transmit simultaneously with Rooney operating in the RF-FD mode. The two data sequences are distinct and two different codebooks are used; 2) Alice and Rooney transmit the data cooperatively using the DF-FD mode. In this case, the data will not be stored at Rooney's queue and, hence, Rooney's queue state does not change in the current time slot; or 3) Rooney operates in the RF-HD mode and either Alice only transmits or Rooney only transmits. Depending on the values of $\mathcal{S}_1$ to $\mathcal{S}_5$, we have the following cases
    \begin{itemize}
    \item If $\mathcal{S}^\star=1$, the RF-FD mode is used.
        \item If $\mathcal{S}^\star=0$ and $\mathcal{S}_3=1$, Alice and Rooney transmit cooperatively using the DF-FD mode.
    \item If $\mathcal{S}^\star=0$ and $\mathcal{S}_3=0$, $\mathcal{S}_4=1$ and $\mathcal{S}_5=0$, Alice transmits to Rooney using the HD mode.
    \item If $\mathcal{S}^\star=0$ and $\mathcal{S}_3=0$, $\mathcal{S}_4=0$ and $\mathcal{S}_5=1$, Rooney transmits to Bob using the HD mode.
      \item If $\mathcal{S}^\star\!=\!0$, $\mathcal{S}_3\!=\!0$, $\mathcal{S}_4\!=\!1$ and $\mathcal{S}_5\!=\!1$, there are two~options:
      \begin{itemize}
        \item If $Q_R=m$ ($m\ge 1$), Alice transmits to Rooney with probability $\alpha_m$ where $\alpha_{\mathcal{Q}}=0$.
            \item If $Q_R=m$, Rooney transmits to Bob using the HD mode with probability $\overline{\alpha_m}=1-\alpha_{m}$.
      \end{itemize}
    \end{itemize}
 If $Q_R=0$, Alice and Rooney have only two transmission options: 1) Alice transmits to Rooney using the HD mode; or 2) Alice and Rooney transmit the data cooperatively using the DF-FD mode. Hence,
            \begin{itemize}
            \item If $\mathcal{S}_3=1$, the DF-FD mode is used.
            \item If $\mathcal{S}_3=0$ and $\mathcal{S}_4=1$, Alice transmits ($\alpha_0=1$).
            \end{itemize}
\underline{\emph{Remark:}} In our proposed scheme, we give priority to the RF-FD mode as it exploits the FD capability, together with an enhanced  secrecy rate. The RF-FD mode cannot be used when $Q_R=0$. However, we can still make use of the FD capability via the DF-FD mode. When FD transmissions are infeasible due to secrecy constraints, we resort to the HD mode.

   \subsection{Rooney's Buffer}
From the described scheme in Section \ref{sec3}, Rooney: 1) neither receives nor transmits data packets, 2) receives one packet, 3) transmits one packet, or 4) receives one packet and transmits one packet in the same time slot. Hence, the Markov chain of Rooney's queue can be modeled as a birth-death process. When the queue is in State $0< n\le \mathcal{Q}$ packets, the probability of the queue state decreasing by $1$ is given by \begin{equation} \small
\begin{split} \small
\label{jkio}
b_n\!&=\!\Pr\{\mathcal{S}^\star\!=\!0,\mathcal{S}_3\!=\!0,\mathcal{S}_4\!=\!1,\mathcal{S}_5\!=\!1\}\overline{\alpha_n} \\& \!+\! \Pr\{\mathcal{S}^\star\!=\!0,\mathcal{S}_3\!=\!0,\mathcal{S}_4\!=\!0,\mathcal{S}_5\!=\!1\}\!=\! k_1 \overline{\alpha_n}\!+\!k_2
\end{split}
\end{equation}
where $k_1=\Pr\{\mathcal{S}^\star\!=\!0,\mathcal{S}_3\!=\!0,\mathcal{S}_4\!=\!1,\mathcal{S}_5\!=\!1\}$, $k_2=\Pr\{\mathcal{S}^\star\!=\!0,\mathcal{S}_3\!=\!0,\mathcal{S}_4\!=\!0,\mathcal{S}_5\!=\!1\}$, and $b_n$ is the transition probability from State $n+1$ to State $n$. When the queue is in State $0\le n <\mathcal{Q}$ packets, the probability of the queue state increasing by $1$ is given by
\begin{equation} \small
\begin{split} \small
\label{aas}
a_n&\!=\!\Pr\{\mathcal{S}^\star\!=\!0,\mathcal{S}_3\!=\!0,\mathcal{S}_4\!=\!1,\mathcal{S}_5\!=\!1\}\alpha_n \\& \!+\!  \Pr\{\mathcal{S}^\star\!=\!0,\mathcal{S}_3\!=\!0,\mathcal{S}_4=1,\mathcal{S}_5=0\} \! =\! k_1 \alpha_n \!+\!k_3
\end{split}
\end{equation}
where $k_3=\Pr\{\mathcal{S}^\star\!=\!0,\mathcal{S}_3\!=\!0,\mathcal{S}_4=1,\mathcal{S}_5=0\}$, and $a_n$ is the transition probability from State $n$ to State $n+1$.

Analyzing the Markov chain of Rooney's queue, the local balance equations are given by
\begin{equation} \small \small
\label{balance}
\small
\small \zeta_n a_n=\zeta_{n+1}b_{n+1}, \ 0\le n \le \mathcal{Q}-1
\end{equation}
where $\zeta_n$ denotes the probability of having $n$ packets at Rooney's queue. Using the balance equations successively, the stationary distribution of $\zeta_n$ is given by
\begin{equation} \small \small
\small
\small \zeta_n=\zeta_0 \prod_{m=0}^{n-1}\frac{a_m}{b_{m+1}} \label{state_probabilities}
\end{equation}
where $\zeta_0=\Big(1+\sum_{n=1}^{\mathcal{Q}}\prod_{m=0}^{n-1} \frac{a_m}{b_{m+1}}\Big)^{-1}$ is obtained using the normalization condition $\sum_{m=0}^{\mathcal{Q}} \zeta_m=1$.

Based on the above description, the average secure end-to-end throughput in packets/slot, denoted by $\mu$, is given by
\begin{equation} \small
\begin{split} \small
\label{e2e}
\mu\!
 &\!=\!\sum_{n=1}^{\mathcal{Q}}b_n \zeta_n\!+\! \Pr\{\mathcal{S}^\star\!=\!1\} \!+\!\Pr\{\!\mathcal{S}^\star\!=\!0,\mathcal{S}_3\!=\!1\!\} \ \ \ \ \ \ \ \\& +\Bigg(\Pr\{\mathcal{S}_3\!=\!1\}-\left(\Pr\{\mathcal{S}^\star\!=\!1\} \!+\!\Pr\{\!\mathcal{S}^\star\!=\!0,\mathcal{S}_3\!=\!1\!\}\right)\Bigg)\zeta_0
\end{split}
\end{equation}
 Our aim is to maximize $\mu$ via optimizing $\{\alpha_m\}_{m=1}^{\mathcal{Q}-1}$. The optimization problem is given by
\begin{equation} \small
\begin{split} \small
\label{e2e22}
\underset{0\le \{\alpha_m\}_{m=1}^{\mathcal{Q}-1} \le 1}{\max:} \ \mu
\end{split}
\end{equation}
The formulation of this optimization problem is nonconvex in $\{\alpha_m\}_{m=1}^{\mathcal{Q}-1}$. Next, we employ a change of variables and convert the problem to a linear program.

Letting $\phi_n=\prod_{m=0}^{n-1}\frac{a_m}{b_{m+1}}$, and substituting with the value of $\zeta_0$, we can rewrite $\sum_{n=1}^{\mathcal{Q}}b_n \zeta_n$ as follows
\begin{equation} \small
\begin{split} \small
\label{e2e22}
\sum_{n=1}^{\mathcal{Q}}b_n \zeta_n
 =\sum_{n=1}^{\mathcal{Q}}((k_1+k_2)\phi_n)\!-\! k_1 \sum_{n=1}^{\mathcal{Q}-1}(\alpha_n\phi_n)
\end{split}
\end{equation}
In the following, we write $\sum_{n=1}^{\mathcal{Q}-1}(\alpha_n\phi_n)$ as a linear function of $\{\phi_n\}_{n=1}^{\mathcal{Q}-1}$. Since $\phi_n=\prod_{m=0}^{n-1}\frac{a_m}{b_{m+1}}$, we have the following relation between $\phi_n$ and $\phi_{n-1}$
\begin{equation} \small
\begin{split} \small
\label{e2e222}
\frac{\phi_n}{\phi_{n-1}}=\frac{a_{n-1}}{b_n}=\frac{k_1 \alpha_{n-1}+k_3}{k_1 \overline{\alpha_n}+k_2}
\end{split}
\end{equation}
Therefore, we get the following recurrence relation between $\phi_n \alpha_n $ and $\phi_{n-1} \alpha_{n-1}$
\begin{equation} \small
\begin{split} \small
{ \phi_n \alpha_n}=-{{\phi_{n-1}} \alpha_{n-1}-\frac{k_3}{k_1}{\phi_{n-1}}}+\frac{k_1+k_2}{k_1} \phi_n
\end{split}
\end{equation}
Letting  $d=\frac{k_3}{k_1}$ and $e=\frac{k_1+k_2}{k_1}$, then
\begin{equation} \small
\begin{split} \small
{ \phi_n \alpha_n}=-{\phi_{n-1}} \alpha_{n-1}-d{\phi_{n-1}}+e \phi_n
\end{split}
\end{equation}

Since $\phi_n=\prod_{m=0}^{n-1}\frac{a_m}{b_{m+1}}$, $\forall n\ge 1$, we have
\begin{equation} \small
\begin{split} \small
\label{initial_valuex}
 {\phi_1}=\frac{a_0}{b_1}=\frac{(k_1\!+\!k_3)}{k_1 \overline{\alpha_1}+k_2}
\end{split}
\end{equation}
Hence,
\begin{eqnarray}
\phi_{1} \alpha_{1} = e \phi_1-\beta, \ \hbox{where} \ \beta=\frac{(k_1\!+\!k_3)}{k_1}
\label{initial_value}
\end{eqnarray}
Starting from (\ref{initial_value}), $\phi_n\alpha_n$ is given by
\begin{equation} \small
\phi_{n} \alpha_{n}= e \phi_n +  \textstyle \sum _{\ell=1}^{n-1} (-1)^{n-\ell} (e+d) \phi_\ell+(-1)^n \beta
\label{general_form}
\end{equation}
Substituting (\ref{general_form}) into \eqref{e2e}, we obtain the following linear-fractional program
\vspace{-1mm}
\begin{eqnarray} \small
\underset{\boldsymbol{\phi}}{\max:} & \frac{\boldsymbol{c}^\top\boldsymbol{\phi}+f}{\boldsymbol{1}^\top\boldsymbol{\phi} +1}\label{LFP}\\
{\rm s.t.} & 0\leq \frac{\boldsymbol{c}_n^\top \boldsymbol{\phi}+g_n}{\phi_n} \leq 1, \forall n \in \{1,2,\cdots,\mathcal{Q}-1\}\nonumber\\
& \small \boldsymbol{c}_{\mathcal{Q}}^\top \boldsymbol{\phi}+g_{\mathcal{Q}} = 0\nonumber
\end{eqnarray}
\noindent where the superscript $\top$ denotes the vector transposition, $\boldsymbol{\phi}=[\phi_{\mathcal{Q}}, \phi_{\mathcal{Q}-1}, \cdots, \phi_2,\phi_1]^\top$, $\boldsymbol{c}=-k_1[-(k_1\!+\!k_2)/k_1,e, -d , e,-d,  \cdots]^\top+(k_1+k_2)[0,1, \cdots,1]^\top$,
\begin{equation} \small {\small
\begin{split}
\!\!\!\!f&\!=\!\left\{\begin{array}{ll}
\!\!\!\! \Pr\{\mathcal{S}_3\!=\!1\}\!-\!\left(\Pr\{\mathcal{S}^\star\!=\!1\} \!+\!\Pr\{\!\mathcal{S}^\star\!=\!0,\mathcal{S}_3\!=\!1\!\}\right), & \mbox{$\mathcal{Q}$ odd}  \\
\!\!\!\! \Pr\{\mathcal{S}_3\!=\!1\}\!-\!\left(\Pr\{\mathcal{S}^\star\!=\!1\} \!+\!\Pr\{\!\mathcal{S}^\star\!=\!0,\mathcal{S}_3\!=\!1\!\}\right)\!-\!\beta, & \mbox{$\mathcal{Q}$ even}
\end{array}\right.\\
\!\!\!\boldsymbol{c}_n&\!=\! [\underbrace{0,\cdots,0}_{\mathcal{Q}-n \text{ zeros}}, e, -(e\!+\!d), (e\!+\!d),-(e\!+\!d), \cdots]^\top,\ g_n \!=\! (-1)^n \beta \notag
\end{split}} \normalsize \!\!\!\!\!
\end{equation}
Note that $\boldsymbol{\phi}$, $\boldsymbol{c}$, and $\boldsymbol{c}_n$ are $\mathcal{Q}$-dimensional column vectors, while $f$ and $g_n$ are scalars. The linear-fractional program in (\ref{LFP}) can be solved efficiently by converting it to a standard linear program \cite[Page 151]{boyed}. Then, we obtain $\{\alpha_n\}_{n=1}^{\mathcal{Q}-1}$~as
{\vspace{-1mm}\begin{eqnarray}
\alpha_n &=& \frac{\boldsymbol{c}_n^\top \boldsymbol{\phi}+g_n}{\phi_n}, \forall n \in \{1,2,\cdots,\mathcal{Q}-1\}
\end{eqnarray}}
 \begin{figure}
  \centering
    \includegraphics[width=1\columnwidth]{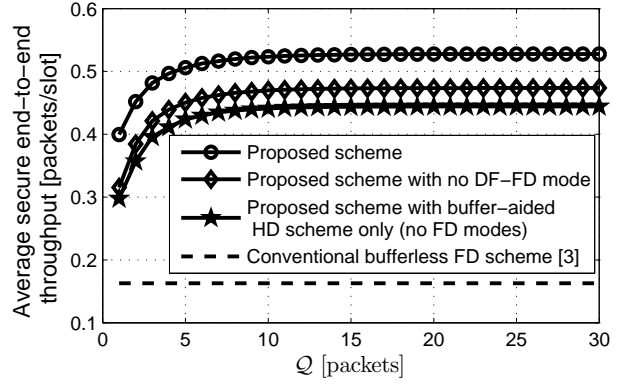}\\
  \caption{Average secure end-to-end throughput versus the maximum buffer size at Rooney, $\mathcal{Q}$. The parameters used to generate the figure are: $\kappa_R=\kappa_B=\kappa_E=\kappa$, $\mathcal{R}_{\rm s}=1$ bits/sec/Hz, $b_{\rm s}=1000$ bits, $W=1$ MHz, $T=1$ msec, $\mathcal{P}_A/(\kappa W)=\mathcal{P}_R/(\kappa W)=10$ dB, and $\sigma_{RR}^2\!=\!0.1$.}\label{fig1}
  \end{figure}
\section{Numerical Simulations and Conclusions}
In this section, we simulate the considered wireless network and show the benefits of our proposed scheme. The fading channels are assumed to be complex circularly-symmetric Gaussian random variables with zero mean and unit variance. In Fig. \ref{fig1}, we plot the average secure end-to-end throughput of our proposed scheme and the conventional bufferless FD scheme. In the conventional bufferless FD scheme, Alice and Rooney cooperatively transmit the data in each time slot using the DF relaying strategy.
 The parameters used to generate Fig. \ref{fig1} are: $\kappa_R=\kappa_B=\kappa_E=\kappa$, $\mathcal{R}_{\rm s}=1$ bits/sec/Hz, $b_{\rm s}=1000$ bits, $W=1$ MHz, $T=1$ msec, $\mathcal{P}_A/(\kappa W)=\mathcal{P}_R/(\kappa W)=10$ dB, and $\sigma_{RR}^2\!=\!0.1$. As shown in Fig. \ref{fig1}, the achievable secrecy rates increase with increasing the buffer size at Rooney. This is expected since increasing the buffer size allows more data transfer to and from Rooney. Moreover, our proposed scheme achieves an average secure end-to-end throughput higher than that achieved by the conventional bufferless FD relaying. The average secure throughput gain is more than $231\%$ when the buffer maximum size is $\mathcal{Q} \ge 4$ packets. We also plot two additional schemes: 1) $\mathbb{S}_1$: our proposed scheme when the DF-FD mode is not used; 2) $\mathbb{S}_2$: our proposed scheme with buffer-aided HD scheme (i.e. no FD modes). As shown in Fig. \ref{fig1}, buffer-aided schemes outperform the bufferless FD scheme. Our proposed scheme with RF-FD and DF-FD modes outperforms $\mathbb{S}_1$ and $\mathbb{S}_2$ with throughput gain of $13\%$ and $20\%$, respectively.
\bibliographystyle{IEEEtran}
\bibliography{IEEEabrv,term_bib}
\end{document}